\documentclass[aps,prl,floatfix,twocolumn,showpacs,superscriptaddress,reprint]{revtex4-1}
\usepackage{amsfonts}
\usepackage{mathrsfs}
\usepackage{amsmath}
\usepackage{color}
\usepackage{graphicx}
\DeclareGraphicsExtensions{.png,.pdf}
\usepackage{bm}
\usepackage{amssymb}
\usepackage{xspace}
\usepackage{epstopdf}
\usepackage{dcolumn}
\usepackage{longtable}
\usepackage{subfigure}
\usepackage{multirow}
\usepackage[colorlinks=true, letterpaper=true, pdfstartview=FitV, linkcolor=blue, citecolor=blue, urlcolor=blue]{hyperref}

\newcommand{\sgn}{\text{sgn}}
\makeatletter

\newcommand{\Rmnum}[1]{\expandafter\@slowromancap\romannumeral #1@}
\makeatother
\begin{document}
\title{
        Zeeman Field-Tuned Transitions for Surface Chern Insulators
       }
\author{Fan Zhang}\email{zhf@sas.upenn.edu}
\affiliation{Department of Physics and Astronomy, University of Pennsylvania, Philadelphia, PA 19104, USA}
\author{Xiao Li}
\affiliation{Department of Physics, The University of Texas at Austin, Austin, Texas 78712, USA}
\author{Ji Feng}
\affiliation{International Center for Quantum Materials, Peking University, Beijing 100871, China}
\author{C. L. Kane}
\affiliation{Department of Physics and Astronomy, University of Pennsylvania, Philadelphia, PA 19104, USA}
\author{E. J. Mele}
\affiliation{Department of Physics and Astronomy, University of Pennsylvania, Philadelphia, PA 19104, USA}
\begin{abstract}
Mirror symmetric surfaces of a topological crystalline insulator host even number of Dirac surface states. A surface Zeeman
field generically gaps these states leading to a quantized anomalous Hall effect. Varying the {\it direction} of
Zeeman field induces transitions between different surface insulating states with any two Chern numbers between $-4$ and $4$.
In the crystal frame the phase boundaries occur for field orientations which are great circles with $(111)$-like normals on a sphere.
\end{abstract}
\maketitle

{\color{cyan}{\indent{\em Introduction.}}}---
The discovery of topological insulators (TI)~\cite{Fu-Kane,Moore-Balents,Roy,RMP1,RMP2} has opened the door for studying
the fascinating interplay between topological order and symmetry breaking
in electronic states of matter.
A boundary of TI hosts gapless states that
exhibit unique properties not realizable in conventional systems.
Importantly, symmetry breaking fields can gap these topological boundary states and provide platforms
for engineering different classes of topological states that are conceptually well known but in practice are hard to design.
Coupling the surface state of a strong (weak) TI to an $s$ ($s_\pm$) wave superconductor breaks the $U(1)$ gauge symmetry
yielding a class D (DIII) topological superconductor (SC) with chiral (helical) Majorana edge states~\cite{D,DIII}.
Zeeman coupling to the surface electron spin breaks time-reversal symmetry producing a quantized anomalous Hall (QAH) insulator~\cite{Haldane} with a chiral edge state~\cite{RMP1,QHZ,film,Zhang-QAH}.
Recently this effect has been observed in a magnetically doped TI thin film~\cite{Xue}.

Chiral edge states around the QAH-insulating TI surface can be produced at a magnetic domain wall~\cite{RMP1},
in a thin film geometry~\cite{QHZ,film}, or more generally at a narrow edge that connects two crystal faces~\cite{Zhang-QAH}.
However, there are two limitations of this approach to creating a surface QAH insulator.
First, an out-of-plane Zeeman field has been proven to be a necessary condition~\cite{warping},
but in many experiments the surface magnetization has an in-plane easy axis.
Secondly, the Chern number can only be $0$ or $1$ in this geometry since the edge connects two gapped surface Dirac spectra
each of which is characterized by a half-integer quantized Hall conductivity.
One might wonder whether it is possible to design a family of QAH insulators with larger Chern numbers,
and whether there is a way to access topological quantum phase transitions between states with different Chern numbers.

\begin{figure}[!b]
\scalebox{0.59}{\includegraphics*{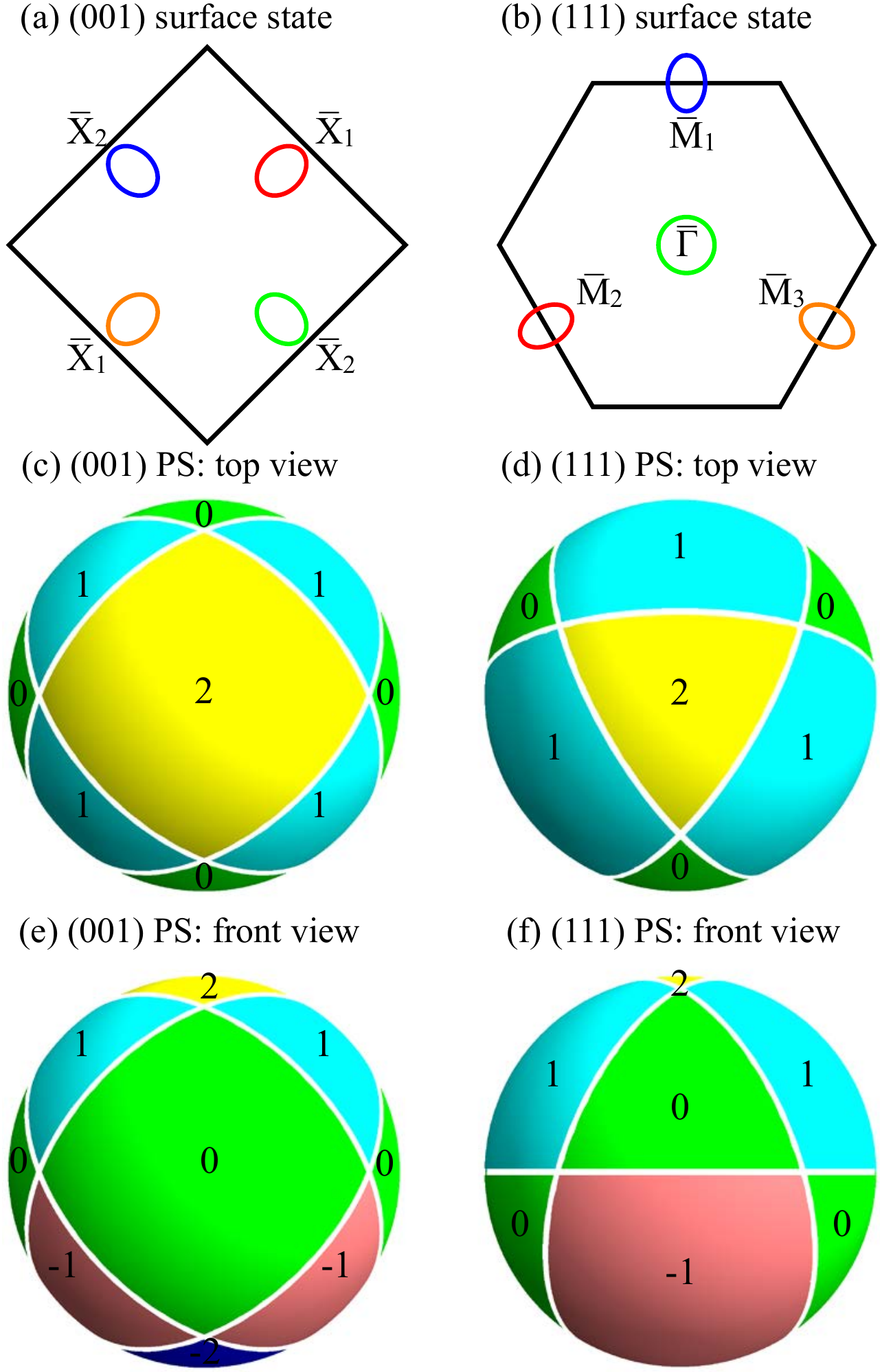}}
\caption{(color online).
(a-b) The protected surface states on $(001)$ and $(111)$ surfaces.
(c-f) The front and top views of $(001)$ and $(111)$ phase spheres (PS).
Each point on a unit sphere denotes a unique direction of Zeeman field.
Different colors represent states with different Chern numbers as labeled.
The four white $(111)$-like great circles are the phase boundaries and the center of yellow region is the local surface normal.
These single-surface PSs are also the building blocks of PSs for two-surface geometries.}
\label{fig1}
\end{figure}

In this Letter, we show that both goals can be readily realized on the mirror symmetric surfaces of $\rm SnTe$, a representative topological crystalline
insulator (TCI)~\cite{TMI,TMI-exp1,TMI-exp2,TMI-exp3,Madhavan,Fiete1}. Though neither completely necessary nor always sufficient,
an {\it ordinary} Zeeman field can break mirror and time-reversal symmetries and thus can generically gap the surface states and lead to a QAH effect.
More remarkably, we find that the {\it direction of the Zeeman field} is a degree of freedom
that allows one to tune the Chern number to any integer between $-4$ and $4$.
We further find that in crystal frame the phase boundaries of QAH insulators with different Chern numbers occur
when the Zeeman field is directed along great circles with $(111)$-like normals. These novel results are plotted in Fig.~\ref{fig1}.

\begin{figure}[t]
(a) Bulk Brillouin zone \qquad (b) Phase boundary distortion
\scalebox{0.45}
{\includegraphics*{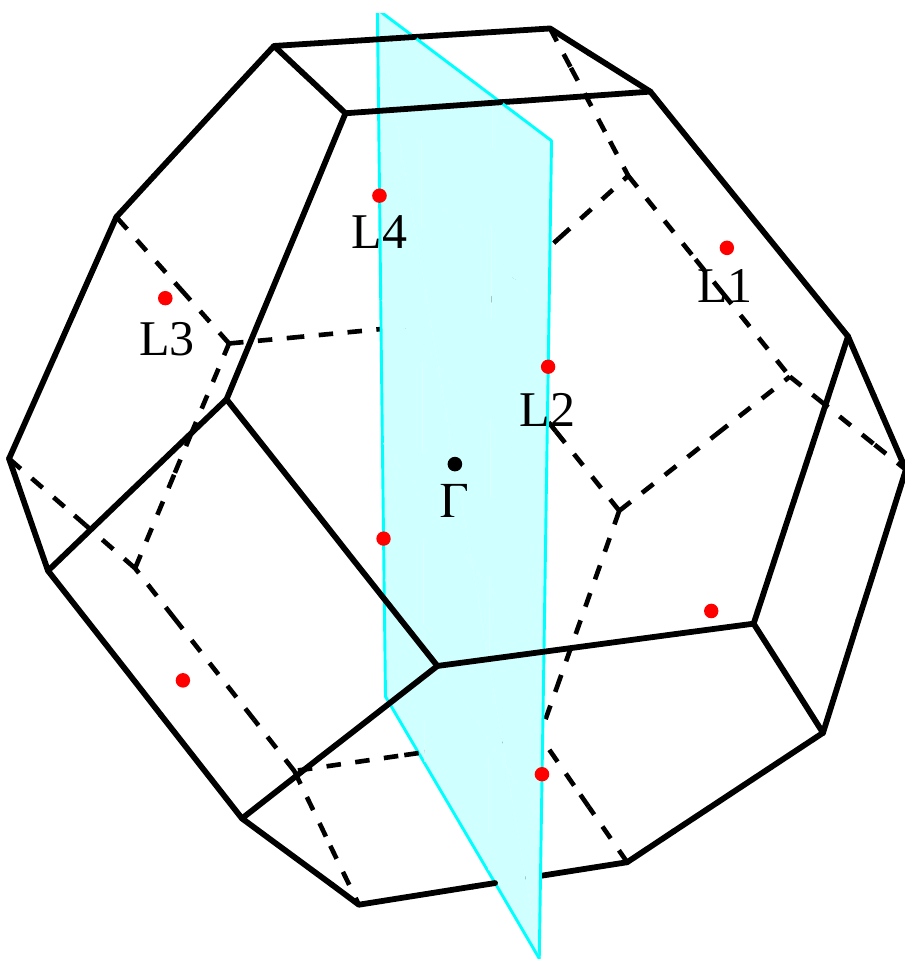}}
\scalebox{0.225}
{\includegraphics*{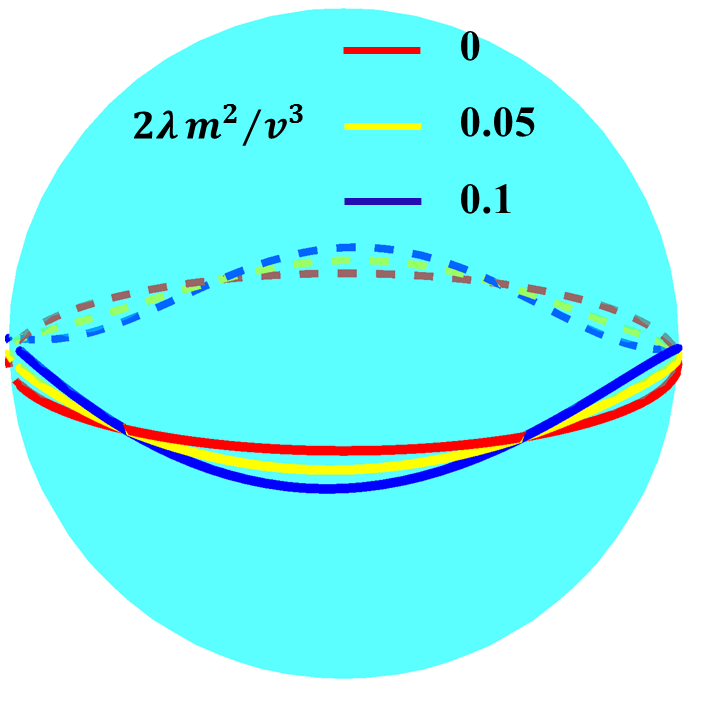}}
\caption{(color online).
(a) The first Brillouin zone, $L$ points, and a $(110)$-like mirror.
(b) Distortion of one of the phase boundaries (red) due to hexagonal warping.
The distortion is invisible when $2\lambda m^2/v^3<0.02$, which is the realistic case.}
\label{fig2}
\end{figure}

{\color{cyan}{\indent{\em Surface states in TCI.}}}---
We start from a description of the symmetries and band inversions in $\rm SnTe$.
$\rm SnTe$ has the rocksalt crystal structure where the two atom types form separate
face-centered cubic lattices that interpenetrate to form a three dimensional checkerboard pattern.
The first Brillouin zone of the crystal structure is a truncated octahedron with six square and eight hexagonal faces,
as shown in Fig.~\ref{fig2}(a). It has been well established that the bands are inverted at four $L$ points~\cite{TMI,1966},
i.e., the centers of four inequivalent hexagonal faces. At each $L$ point, there is an inversion symmetry
($\mathcal{P}=\sigma_z$), a threefold rotational symmetry ($\mathcal{C}_3$) with respect to $\Gamma L$, and three $(110)$-like
mirror symmetries ($\mathcal{M}_2=-is_2$). Therefore, the $k\cdot p$ Hamiltonian near a given $L$ point to linear order in
momentum has a unique form
\begin{eqnarray}\label{HL}
\mathcal{H}_L=v(k_2s_1-k_1s_2)\sigma_x+v_3k_3\sigma_y+\Delta\sigma_z\,,
\end{eqnarray}
where $\bm k$ is described in the {\it local} frame with $\hat{k}_3=\Gamma L$,
$\Delta$ is the bulk gap at $L$ point,
$s_3=\pm 1$ labels the angular momentum $j_z=\pm1/2$ along $\Gamma L$,
and $\sigma_z=\pm1$ corresponds to the p-orbitals on the cation ($\rm Sn$) and anion ($\rm Te$).

Interestingly, this theory at the $L$ point is similar to the theory at the $\Gamma$ point of $\rm Bi_2Se_3$, and indeed both
are determined by the $D_{3d}$ point group symmetries that leave the two points invariant. However, their topological
mechanisms are completely different. The single band inversion in $\rm Bi_2Se_3$ guarantees that it is a
strong TI protected by time-reversal symmetry, whereas in the same theory the four band inversions occurring in a common plane,
as seen in Fig.~\ref{fig2}(a), renders $\rm SnTe$ a trivial insulator. However, a {\it mirror} Chern number
exists in both cases and qualifies both as TCIs. In the mirror-invariant plane with $k_2=0$, $\mathcal{H}_L$ decomposes into
\begin{eqnarray}
\mathcal{H}_{\pm}=\pm vk_1\sigma_x+v_3k_3\sigma_y+\Delta\sigma_z\,,
\end{eqnarray}
each of which describes a two dimensional massive Dirac fermion in a mirror subspaces $\mathcal{M}_2=\pm i$.
The band inversion ($\Delta$ switching sign) changes
the Chern number by $\pm 1$ for $H_{\pm}$ at the interface to vacuum or $\rm PbTe$~\cite{TMI}.
As a result, a gapless Dirac surface state appears on any surface
that respects the mirror symmetry~\cite{TMI,TMI-exp1,TMI-exp2,TMI-exp3}. Note that there are six inequivalent $(110)$-like mirrors and each mirror contains two of
the four inequivalent $L$ points.

We now implement the topological boundary condition~\cite{Zhang-TISS} to derive the long wavelength theory for the
surface state due to the band inversion at a given $L$ point. The mass $\Delta$ is positive and large in the exterior of TCI
whereas in the interior is inverted, i.e., negative and finite. This boundary problem can be solved yielding a surface state
described by
\begin{eqnarray}
\label{HS}
\mathcal{H}_{sf}(\theta_s)=v_x k_x \bar{s}_y-v k_y \bar{s}_x\,,
\end{eqnarray}
where $k_y=k_2$, $\hat{k}_z$ is the {\it local} surface normal, and $k_x$ is rotated from $k_1$ by $\theta_s$ along $\hat{k}_2$
such that the surface preserves the $\mathcal{M}_2$ mirror symmetry;
$v_x=vv_3/v_z$ and $v_z=\sqrt{(v_{\rm 3}\cos\theta_s)^2+(v\sin\theta_s)^2}$.
The surface state is the negative eigenstate of $\bar{\sigma}_x$
and its chiral (positive) counterpart is localized on the opposite surface.
Thus any surface state only inherits {\it half} of the bulk degrees of freedom
and its corresponding pseudospins $\bar{\bm\sigma}$ and $\bar{\bm s}$ read
\begin{eqnarray}
\label{structure}
\bar{\bm\sigma}&=\{\alpha\sigma_1+\beta s_2\sigma_2,\alpha\sigma_2-\beta s_2\sigma_1,\sigma_3\}\,,&\\
\bar{\bm s}&=\{\alpha s_1-\beta s_3\sigma_3,s_2,\alpha s_3+\beta s_1\sigma_3\}\,,&
\end{eqnarray}
where $\alpha=v_3\cos\theta_s/v_z$, $\beta=v\sin\theta_s/v_z$, and
$[\tau^a_i,\tau^b_j]=2i\delta^{ab}\epsilon_{ijk}\tau_k^a$ with ${\bm\tau}^a$ being $\bar{\bm s}$ or $\bar{\bm \sigma}$.
Under time-reversal, inversion, and mirror reflection, $\bar{\bm s}$ and $\bar{\bm \sigma}$ behave in the same manner as ${\bm
s}$ and $\bm \sigma$. The former becomes latter when $\theta_s=0$.
The surface state pseudospin ($\bar{\bm s}$) texture is topologically equivalent to that of a helical metal,
e.g., the $(111)$ $\bar{\Gamma}$ surface state. However, on each surface the {\em spin} ($\bm s$) texture~\cite{Zhang-TISS} is
unique. This implies that the response to Zeeman field is surface-dependent.

{\color{cyan}{\indent{\em Hall conductivity.}}}--- To induce a QAH effect, a surface perturbation must break time-reversal
symmetry and dominate other gap-opening mechanisms not responsible for Hall effects, e.g., inter-surface hybridization.
This is readily satisfied by an {\it ordinary} Zeeman coupling term  ${\bm m}\cdot{\bm s}$, which can be introduced by magnetic
doping~\cite{Xue,dop1,dop2,dop3} or proximity coupling to a ferromagnetic insulating layer. Peculiarly, the Dirac surface state
inherits only half of the bulk degrees of freedom, i.e., $\bar{\sigma}_x=-1$ as derived earlier. Thus only the part of ${\bm
m}\cdot{\bm s}$ that commutes with $\bar{\sigma}_x$ leads to an essential surface disturbance.
Conversely, any part that anticommutes with $\bar{\sigma}_x$ simply couples the opposite surfaces which is exponentially weak on the scale of penetration length. In light of this
decomposition rule we obtain the partial Hall conductivity
\begin{eqnarray}\label{CN}
\sigma_H=\frac{1}{2}\sgn(\frac{\pi}{2}-\theta_{s})\sgn(\hat{m}\cdot\hat{k}_3)
\end{eqnarray}
in unit of $e^2/h$. Evidently $\sigma_H$ switches signs when $\bm m$ is rotated across the plane perpendicular to $\hat{k}_3$.
The existence of such a plane indicates that breaking the mirror symmetry is required to gap the surface states. At linear
order in $\bm k$ the $\mathcal{C}_3$ symmetry upgrades to a continuous rotational symmetry and there are infinite number of
mirrors (instead of just three) perpendicular to the critical plane. We shall postpone analyzing the negligible distortion of this
critical plane due to higher order corrections. Nevertheless, if a unit sphere is used to represent the direction of Zeeman
field $\hat{m}=(\sin\theta\cos\phi,\sin\theta\sin\phi,\cos\theta)$, for a surface state arising from the band inversion at
$(111)$-$L$ point, the phase boundary where $\sigma_H$ switches signs is the $(111)$ great circle in the crystal frame.

For strong TIs Eq.~(\ref{CN}) is the half-quantized Hall anomaly~\cite{Fu-Kane,QHZ,Zhang-QAH}. To resolve this anomaly
requires combining the behaviors of two Dirac surface states, in the earlier mentioned two-surface geometry, such that
their sum or difference in their Hall conductivity is an integer. In sharp contrast, this anomaly is automatically resolved for
the most interesting surfaces of TCI~\cite{anomaly}, as each of these hosts {\it even} number of Dirac surface states.
One mirror symmetric surface is meaningful in transport as it can be isolated. Breaking the mirror symmetry on the side surfaces
while leaving them time-reversal invariant, gaps out the side surface states without generating any Hall contribution.

{\color{cyan}{\indent{\em Phase Spheres.}}}--- We now consider the QAH effect on the $(001)$ surface, where the four surface
states near $\bar{X}$ points are related by a $\mathcal{C}_4$ symmetry, as sketched in Fig.~\ref{fig1}(a). When the Zeeman
field is perpendicular to the surface, the four surface states are all gapped and each contributes $1/2$ in $\sigma_H$ with the
same sign, as required by the $\mathcal{C}_4$ symmetry. In total this state has $\sigma_H=2$. When $\hat{m}$ is rotated and
crosses one of the four $(111)$-like great circles, i.e. the phase boundaries, one of the four surface state undergoes a gap
closing and reopening and its $\sigma_H$ switches sign signalling a transition to the state with total $\sigma_H=1$.
If $\hat{m}$ instead crosses an intersection point of {\it two} circles, two surface states will undergo the above mentioned phase transition and
the total $\sigma_H$ will vanish. $\sigma_H$ changes its sign if $\hat{m}$ is reversed, and the whole phase sphere can thus be determined,
as shown in Fig.~\ref{fig1}(c) and (e). Note that without crossing any great circle, rotating the Zeeman-field
cannot change $\sigma_H$, as the surface state remains gapped and the ground states are adiabatically connected.

On the $(111)$ surface, the three surface states at inequivalent $\bar{M}$ points are related by a $\mathcal{C}_3$ symmetry, as
sketched in Fig.~\ref{fig1}(b). Although the Dirac points are independent at $\bar{M}$ and
$\bar{\Gamma}$ points, a symmetry-allowed surface potential~\cite{Zhang-TISS} can tune them to the same energy,
as we will assume implicitly. A perpendicular Zeeman field gaps all the surface states
and it is anticipated that, as required by the $\mathcal{C}_3$ symmetry, the three surface states at the $M$ points contribute
equally to the total $\sigma_H$. However, it is nontrivial to determine whether $\sigma_H$ of the surface state at $\bar{\Gamma}$ point
would have the same sign. It turns out that the projection angles of bulk band inversions for $\bar{M}$ and $\bar{\Gamma}$
surface states are $\theta_s(\bar{M})<\pi/2$~\cite{M-angle} and $\theta_s(\bar{\Gamma})=0$, respectively. Consequently,
evaluating Eq.~(\ref{CN}) reveals that the $M$ and $\Gamma$ surface states have the same sign in $\sigma_H$, i.e., $\sigma_H=2$
in total. As we vary the direction of Zeeman field $\hat{m}$, the total $\sigma_H$ remains $2$  unless $\hat{m}$ crosses one of the three
$(111)$-like great circles or an intersection point of two circles. In the former (latter) case, the total $\sigma_H$ becomes
$1$ ($0$) as one (two) of the $\bar{M}$ surface states switches signs in $\sigma_H$. Noting that $\sigma_H$
changes sign if $\hat{m}$ is reversed, we can complete the phase sphere for $(111)$ surface states, as plotted in
Fig.~\ref{fig1}(d) and (f).

In the phase spheres, for the $(001)$ surface there is a $\mathcal{C}_4$ symmetry along the local surface normal whereas for
the $(111)$ surface the symmetry is $\mathcal{C}_3$. This implies that the observed QAH effects should have a $2\pi/3$ or
$\pi/2$ periodicity if we rotate the Zeeman field around the surface normal. Note that we can also consider the $(110)$
surface. Such a surface respects one mirror symmetry and thus hosts one pair of surface states. As a result, the total $\sigma_H$ is only
allowed to be $0$ or $\pm1$ by varying the field direction, with four regions separated by two great circles in the
corresponding phase sphere (not shown).

We can further consider thin-film geometries. When the top and bottom surfaces share the same Zeeman field, their contributions
to $\sigma_H$ are the same in the local frame of one of the two surfaces. Therefore, the Chern number in the thin-film phase
sphere will be doubled compared with those shown in Fig.~\ref{fig1} for one single surface. If the field direction on either
surface can be independently controlled, then QAH states with odd Chern numbers between $-4$ and $4$ can be realized. More
generally, we can consider a narrow edge connecting two mirror symmetric surfaces, or two surfaces weakly breaking mirror
symmetry~\cite{weakly}. Applying Zeeman fields leads to
\begin{eqnarray}
\sigma_H^T=\sigma_H(\hat{m}_T)-\sigma_H(\hat{m}_B)\in\{0,\pm1,\pm2,\pm3,\pm4\}\,,
\end{eqnarray}
where $\hat{m}_{T/B}$ is the field direction in the top or bottom surface and either $\sigma_H$ is evaluated in its local frame using Eq.~(\ref{CN}).
Specially for the $(111)$ thin film, $\bar{\Gamma}$ and $\bar{M}$ surface states are not related by any symmetry,
therefore when the film is thin enough to produce inter-surface hybridization,
the two gap-opening mechanisms can dominate at different valleys leading to a richer phase diagram.

Besides transport, QAH effects are also observable in optical experiments~\cite{Tse,Maciejok,Armitage}. Consider a linearly polarized light
normally incident on a TCI thin film.  In the low-frequency limit, the transmitted light exhibits a universal Faraday rotation
with $\theta_F=\arctan(\alpha\sigma_H^T)$ whereas the reflected light exhibits a giant universal Kerr rotation
$\theta_K=\arctan[1/(\alpha\sigma_H^T)]$, which are tunable by the field direction. Here $\alpha=1/137$ and $\sigma_H^T$ is
dimensionless.

{\color{cyan}{\indent{\em Distortion of phase boundaries.}}}--- One may wonder whether the higher order terms neglected in
Eq.~(\ref{HL}) would change the phase boundaries dramatically. We find that such distortions exist, but are negligibly small.
For example, consider the case for the surface state at $\bar{\Gamma}$ point on $(111)$ surface ($\theta_s=0$), which can be described by~\cite{Fu-warping}
\begin{eqnarray}\label{HW}
\mathcal{H}_w=v(k_2s_1-k_1s_2)+\lambda(k_+^3+k_-^3)s_3+{\bm m}\cdot{\bm s}\,,
\end{eqnarray}
where $k_\pm=k_2\pm ik_1$ so that $k_+^3+k_-^3$ is odd under $\mathcal{M}_2$ mirror reflection;
the $\lambda$ term is the most important hexagonal warping correction that is restricted by $\mathcal{C}_3$ symmetry.
The Zeeman coupling gaps the $\bar{\Gamma}$ surface state except on a critical curve, i.e., the phase boundary, determined by
\begin{eqnarray}
2\lambda m^2\sin^3\theta\cos3\phi=v^3\cos\theta\,.
\end{eqnarray}
When $\lambda=0$ the phase boundary is the $\theta=\pi/2$ equator as we derived in Eq.~(\ref{CN}). When $\lambda$ is finite, the phase
boundary not only depends on the direction of Zeeman field but also on its strength. As plotted in Fig.~\ref{fig2}(b), when
$\lambda$ is turned on the phase boundary is a closed curve with only six points~\cite{Henk} symmetrically fixed on the
undistorted equator. When pointing to these directions, the Zeeman field is not able to gap the $\bar{\Gamma}$ surface state as
it does not break the $\mathcal{M}_2$ mirror symmetry. Our analysis suggests that even an in-plane field on the $(111)$
surface~\cite{in-plane}, can in principle lead to a QAH effect with a $2\pi/3$ periodicity in the field orientation.

Typically in TIs, $v\sim1-5$~eV$\cdot${\AA}, $\lambda\sim50-100$~eV$\cdot${\AA}$^3$, and $m\sim1-20$~meV. Thus the distortion
ratio $2\lambda m^2/v^3$ is smaller than $10^{-4}m^2$ with $m$ in unit of meV. Nevertheless, as clearly shown in
Fig.~\ref{fig2}(b), such distortions are small and the phase boundary is well approximated by the great circle. The small role
played by high-order warping terms near Dirac points reflects the fact that both the coherence and penetrate lengths of
surface states have a range comparable to $\Delta/\hbar v$, which is typically much larger than the lattice constant. Similar
analysis holds in general for any other surface state, though the analysis is more complicated than for this case.

{\color{cyan}{\indent{\em Discussions.}}}--- Realizing a QAH effect has been explored in a few realistic systems including
strong TIs~\cite{QHZ,film,Zhang-QAH,JH}, graphene few-layers~\cite{Zhang-SQH,Qiao}, and transition metal oxides~\cite{DX,Fiete,FW}.
Recently Chen, Gilbert, and Bernevig suggested to tune the Chern number in TCI $(001)$
thin films, by combining perpendicular Zeeman and electric fields, strain effects, and interlayer couplings~\cite{Bernevig}.
Our scheme of tuning Chern number simply by rotating Zeeman field constitutes a few critical advances.
(i) The induced Chern number can be large and even tunable in a fairly simple but still powerful way,
without using any strain effect~\cite{Bernevig} or a giant Zeeman field~\cite{JH}.
(ii) Unlike the case for a strong TI where a two-surface geometry is required,
a single mirror symmetric surface of TCI is sufficient for observing a QAH effect.
(iii) Almost every field direction can potentially induce a transition.
Even an in-plane magnetization may naturally lead to a QAH effect,
as a result of the trigonal or tetragonal warping effect. This may be important in practice, as in many
cases the magnetization tends to have an in-plane easy axis. (iv) The Hall currents carried by chiral edge states are
dissipationless. Through an external control of the direction of Zeeman field, our scheme may be useful for designing low-power
digital electronics, where signals can be transmitted without degradation due to noise. (v) The Kerr (Faraday) effect requires
a reflected (transmitted) light to undergo a precise rotation in polarization, determined by the Chern number and tunable by
the field direction. This universal effect may be useful in the optics that requires high precision and topological robustness.

For a dual-gated TCI thin-film, the surface states would be trivially gapped if tunneling induced hybridization dominates,
whereas exciton condensation~\cite{Min,Franz} would occur if the two surfaces are strongly correlated by electron interactions.
In the presence of a magnetic field, there is an intriguing competition between the above $\sigma_H=0$ states
and the QAH states. This order competing can be tested in experiment with a tilted magnetic field.
Moreover, coupling our QAH thin film with an $s$-wave SC gives a chiral topological SC,
in which the number of Majorana edge states can be tuned by rotating the Zeeman field and adjusting the chemical potential.

Our study provides a framework to understand the surface states and their magnetization in TCIs.
Its resulting symmetry breaking near multiple Dirac points taken together leads to QAH effects
with a fairly simple but still powerful knob to tune the Chern number.
In this Letter we have assumed an {\it ordinary} Zeeman field which couples the angular momenta
but not the orbitals represented by $\sigma_z=\pm$ here.
If it turns out in experiments the $g$-factor shows strong orbital dependence,
it would be necessary to implement the decomposition rule to analyze the Zeeman effects,
which may have strong surface-dependence and lead to more exotic physic awaiting to be discovered.

\bibliographystyle{apsrev4-1}

\begin{thebibliography}{100}
\bibitem{warping}
 This is true only when we neglect the hexagonal warping, which is indeed negligible as we shall show later.
\bibitem{anomaly}
 This anomaly still exists on the $(11\bar{2})$ surface where $\theta=\pi/2$ in Eq.~(\ref{CN}). There is a pair of surface states,
 but an ordinary Zeeman coupling is not able to gap the one that arises from the band inversion at $(111)$-$L$ point.
\bibitem{M-angle}
 This is the angle between the surface normals of $(111)$ and $(11\bar{1})$ surfaces.
\bibitem{weakly}
 This almost requires the Miller index to be very close to either $(001)$ or $(11n)$.
\bibitem{in-plane}
 This is also true for a $\rm Bi_2Se_3$ thin film with natural surfaces. $\rm Bi_2Se_3$ is not only a strong TI but also a TCI.
\end{thebibliography}

\end{document}